\documentclass[preprint,showpacs,preprintnumbers,amsmath,amssymb]{revtex4}

\usepackage{graphicx,subfigure}
\usepackage{dcolumn}
\usepackage{bm}
\usepackage{psfrag}

 \newcommand{\f}[2]{\frac{#1}{#2}}
 \newcommand{\go}{\rightarrow}
 \newcommand{\p}{\phi}
 \newcommand{\g}{\gamma}
 \newcommand{\ep}{\epsilon}
 
 \newcommand{\E}[1]{10^{#1}\,}
 \newcommand{\e}[1]{10^{- #1}\,}
\begin{document}

\bibliographystyle{apsrev}

\preprint{UAB-FT-579}

\title{Evading Astrophysical Constraints on Axion-Like Particles} 

\author{Eduard Mass{\'o}}
\author{Javier Redondo}
\affiliation{Grup de F{\'\i}sica Te{\`o}rica and Institut
de F{\'\i}sica d'Altes
Energies\\Universitat Aut{\`o}noma de Barcelona\\
08193 Bellaterra, Barcelona, Spain}


\date{\today}

\begin{abstract}
Stellar energy loss arguments lead to strong constraints
on the coupling $\phi \gamma \gamma$
of a light axion-like particle  to two photons.
Helioscopes, like CAST, are able to put competitive bounds.
The PVLAS experiment has recently observed a rotation of the
polarization of a laser propagating in a magnetic field that
can be interpreted as the effect of a quite strong
$\phi \gamma \gamma$ coupling. We present scenarios where
the astrophysical and CAST bounds can be evaded,
and we show that the PVLAS result can be accomodated in one
of the models, provided the new physics scale is
at very low energies.
\end{abstract}

\maketitle



\section{Introduction}

An pseudoscalar axion-like particle $\phi$ coupled to photons
\begin{equation}
{\cal
L}_{\phi\gamma\gamma}=\f{1}{4M}F^{\mu\nu}\widetilde{F}_{\mu\nu}\p
\label{eq:1a}
\end{equation}
($F^{\mu\nu}$ is the electromagnetic field tensor,
$\widetilde{F}_{\mu\nu}$ its dual, and $\phi$ the axion-like
field) would be able to transform into photons when electromagnetic
fields are present
\begin{equation}
{\cal
L}_{\phi\gamma\gamma}=\f{1}{M}\overrightarrow{E}\cdot\overrightarrow{B}\,
\p\label{eq:1b}
\end{equation}
This is completely analogous to the well known Primakoff effect that
involves the $\pi^0\gamma\gamma$ coupling.
In this paper we will be interested in the case that $\p$ is very light
since then a number of interesting effects may happen.

When the $\p$ mass $m_\p<T_c$, with $T_c \simeq 1 - 10$ keV the
typical temperature of stellar cores as those of the Sun or
horizontal-branch stars, $\p$ particles are produced
by the Primakoff-like effect due to
the interaction (\ref{eq:1a},\ref{eq:1b}). If one further
assumes that the produced $\p$ flux escapes freely from the star
and thus constitutes a non-standard channel of energy-loss,
the strength of the interaction (\ref{eq:1a},\ref{eq:1b}) can be
bounded using observational data on stellar evolution time
scales \cite{Raffelt:1999tx}. For the Sun, one has the limit
\begin{equation}
M \gtrsim 4 \times \E{8}\ {\rm GeV} \label{sun}
\end{equation}
There is general agreement that these arguments applied to globular
clusters lead to an even stronger bound
\begin{equation}
M \gtrsim \E{10}\ {\rm GeV} \label{eq:2}
\end{equation}
(Here and thereafter, we understand that the bounds are on the
absolute value of $M$.)

In the allowed range for $M$, axion-like particles are still
produced in the Sun and a calculable flux reaches the Earth
\cite{flux}. A proposal to detect this
flux was given in a pioneer paper by Sikivie
\cite{Sikivie:1983ip}. The idea is that the
interaction (\ref{eq:1a},\ref{eq:1b}) allows solar $\phi$'s
to transform back into X-ray
photons in a cavity with an external magnetic field. Such
helioscopes have been built and limits on $M$
have been obtained from the non-observation of
this inverse Primakoff process in the cavity
\cite{helioscope,Andriamonje:2004hi}.
The last result,
from the CERN Axion Solar Telescope (CAST) collaboration
\cite{Andriamonje:2004hi}, is
\begin{equation}
M > 0.9 \times\E{10}  \ {\rm GeV}   \label{eq:3}
\end{equation}
($95\%$CL) comparable to the bound based on the stellar energy loss
arguments, eq.(\ref{eq:2}). The strong
limit (\ref{eq:3}) can be established when a coherent $\p\go\g$ signal is
expected, which happens for  $m_\p \lesssim 0.02$
eV. The CAST prospects \cite{Andriamonje:2004hi} are to further
improve (\ref{eq:3}) and to extend the results to masses $m_\p$
up to 1 eV.

The research we present in this letter has been motivated by the
observation  of a rotation of the polarization
plane of light propagating through a transverse, static, magnetic field
by the PVLAS collaboration \cite{pvlas}.
A possible interpretation of this result is the existence of a light axion-like
particle $\p$ coupled to two  photons \cite{Maiani:1986md}.
However, if interpreted this way the scale appearing in
(\ref{eq:1a},\ref{eq:1b}) must be
\begin{equation}
M \simeq 4 \times\E{5}  \ {\rm GeV}   \label{eq:pvlas}
\end{equation}
It results in such a strong coupling that it is in contradiction
with the bounds (\ref{eq:2}) and (\ref{eq:3}).
Yet, it is consistent with
the bounds coming from particle physics experiments  \cite{Masso:1997ru}.

Let us stress that if there exists a particle with the coupling $M^{-1}\simeq
2.5 \times\E{-6}  \, {\rm GeV}^{-1}$ as given in (\ref{eq:pvlas}),
 it definitely cannot be the standard QCD
axion. The naming "axion-like" we use in the paper refers to the particle being
very light and to its pseudoscalar nature, reflected in the form of the  interaction
(\ref{eq:1a}).

Since at present there is no an alternative explanation of the PVLAS
data, we are faced to the challenge of finding a consistent model
that could explain the constraints
(\ref{eq:2}),  (\ref{eq:3}), and (\ref{eq:pvlas}) in terms of a light particle coupled
to two photons.
The route we have followed has been investigating ways to evade the
astrophysical bounds.
 We have worked out two
 possibilities that could solve the problem. The first is that
$\phi$-particles are indeed produced in the Sun but that they interact
so strongly
 that are trapped by the solar medium. Then, the energy of the
 emitted $\p$-particles is much lower than in the usual free-streaming
 regime and thus the CAST telescope is not able to detect them.
In Section \ref{trapping} we propose a simple model with paraphotons
that provides a way $\p$-particles are trapped. However, we
will see that it leads to too strong photon-paraphoton interactions that are
not consistent with other observations.
Even having this problem, we present the
model because, first of all, it remains to be seen
whether a sophistication of these ideas may lead to a consistent model.
Second, some of the issues we are faced are helpful in Section
\ref{suppression}.
The second possibility we examine is that
 the Primakoff process is suppressed when occurring
in a stellar medium.
 Then, there would be far less $\p$-particles emitted than expected.
We discuss in Section \ref{suppression} how a composite $\phi$ and
the corresponding form factor
of the $\p\g\g$ vertex could be responsible
for such a suppression of the
$\p$-flux. Finally, we present our conclusions and additional comments
in Section \ref{conclusions}.

 \section{Trapping regime} \label{trapping}

 Let us start analyzing the possibility that $\p$-particles are produced
in the solar core but that interact so strongly with the medium
that their fate is analogous to what happens to the stellar photons, namely,
they abandon the Sun after a lot of interactions, having
followed a random walk path. In this trapping regime, local
thermodynamic equilibrium applies and $\p$ would contribute to the
radiative energy transfer. The total opacity, including the exotic contribution,
\begin{equation}
k_{total}^{-1}=k_\g^{-1}+k_\p^{-1} \nonumber
\end{equation}
should not be much different from the standard solar opacity $k_\g^{-1}
\approx 1$  g/cm$^2$, if we do not want to ruin the standard solar model. Specifically, one imposes  \cite{samuel}
\begin{equation}
k_\p^{-1} \lesssim 1 \f{\rm g}{{\rm cm}^2} \label{eq:5}
\end{equation}

The key point of course is to try to implement this possibility
within a particle physics model.
The scenario we shall examine assumes paraphotons
provide the trapping interaction.
These vector particles were proposed by Okun in \cite{Okun:1982xi} (see also
\cite{Georgi:1983sy}) and further developed by Holdom in
\cite{Holdom}. The basic idea is a
modification of QED that consists in adding an extra $U(1)$ abelian
gauge symmetry. If $j_\mu$ is the electromagnetic current
involving charged particles $j_\mu \sim \bar{e}\g_\mu e + ...$ we
start with the lagrangian
\begin{equation}
{\cal L}_0 =-\f{1}{4}F_1^{\mu\nu}F_{1\mu\nu}+ e_1j_\mu A_1^\mu
\end{equation}
This lagrangian has a $U(1)$ gauge symmetry group, and would be the
photon part of the QED lagrangian. The paraphoton
model assumes two groups $U(1)_1\times U(1)_2$ as the gauge symmetry, so that
one has two gauge fields $A_1$ and $A_2$.

In the line of \cite{Holdom} we will
assume that there are very massive particles carrying charges
under both $U_1$ and $U_2$ groups.
At low energies, these massive particles running in loops can be
integrated out leaving the lagrangian
\begin{eqnarray}
&{\cal L}
=-\f{1}{4}(1+2\ep_{11})F_1^{\mu\nu}F_{1\mu\nu}-\f{1}{4}(1+2\ep_{22})F_2^{\mu\nu}F_{2\mu\nu} \nonumber\\
&+\f{1}{2}\ep_{12}F_1^{\mu\nu}F_{2\mu\nu}+ e_1j_\mu A_1^\mu
\label{eq:6}
\end{eqnarray}
The parameter $\ep_{12}$ is the induced mixing in the kinetic
terms, and $\ep_{11}$ and $\ep_{22}$ are also modifications due to these
loops. At first order in the small $\ep$-parameters,
we define new fields that
diagonalize and normalize  the kinetic terms,
\begin{eqnarray}
&A_\mu = (1+\ep_{11})A_{1\mu} \label{eq:7} \\ \nonumber &A'_\mu =
(1+\ep_{22})A_{2\mu} - \ep_{12}A_{1\mu}
\end{eqnarray}
We end up with the photon $A_\mu$ coupled to charged particles,  with
$e=e_1(1-\ep_{11})$, and with the paraphoton $A'_\mu$
\begin{equation}
{\cal L}
=-\f{1}{4}F^{\mu\nu}F_{\mu\nu}-\f{1}{4}F'^{\mu\nu}F'_{\mu\nu}+ej_\mu
A^\mu \label{eq:8}
\end{equation}

Different authors have added some physics to (\ref{eq:6}) and (\ref{eq:8}) so that phenomenological consequences arise.
In \cite{Okun:1982xi} and \cite{Georgi:1983sy} the effects of a
paraphoton mass were discussed. In
\cite{Holdom}, it was shown that the
existence of light particles having $U_2$ charge leads to these
particles having an electric charge of size $\ep e$. In
\cite{Foot:2000vy} the paraphoton was identified with a mirror photon and
some implications were analyzed. The most recent work \cite{Dobrescu:2004wz}
considers higher-order operators to describe the interaction of the
paraphoton with matter.

What we propose is to add to ${\cal L}$ in (\ref{eq:6}) the interaction
\begin{eqnarray}
{\cal L}_{\p\g_2\g_2}=\f{1}{4
M_2}F_2^{\mu\nu}\widetilde{F}_{2\mu\nu}\p
\end{eqnarray}
with $M_2$ a low energy scale. The axion-like particle is
therefore strongly coupled to the $U_2$ gauge boson. After
diagonalizing (\ref{eq:7}) we get a strong coupling of $\p$ to
paraphotons
\begin{eqnarray}
{\cal L}_{\p\g'\g'}=\f{1}{4
M_2}F'^{\mu\nu}\widetilde{F}'_{\mu\nu}\p
\end{eqnarray}
a weaker coupling with a mixed term
\begin{eqnarray}
{\cal L}_{\p\g\g'}=\f{\ep_{12}}{2
M_2}F^{\mu\nu}\widetilde{F}'_{\mu\nu}\p
\end{eqnarray}
and finally we get a term that couples $\p$ to photons, i.e., an
interaction as in (\ref{eq:1a}) with the identification
\begin{eqnarray}
\f{1}{M}=\f{\ep_{12}^2}{M_2}\label{eq:9}
\end{eqnarray}
with $M$ giving the strength (\ref{eq:pvlas}).

We now have the necessary ingredients to have a large opacity of
$\p$ in the solar medium. The dominant contribution to the opacity
comes from the process
\begin{eqnarray}
\p\g\go\p\g'\label{A}
\end{eqnarray}
where a virtual $\g '$ is exchanged.
The secondary paraphotons are further scattered
\begin{eqnarray}
\g'\g\go\g'\g'\label{B}
\end{eqnarray}
where now a $\p$ is exchanged.
In both reactions, (\ref{A}) and (\ref{B}), the initial $\g$ is
of course from the stellar plasma.

Having exposed the main idea, we proceed to the calculation
of the opacity, where
we shall content ourselves with order of magnitude estimates.
The head-on collision in (\ref{A}) has a total cross-section
\begin{eqnarray}
\sigma_\p=  \f{5}{384\pi} \left(\f{\ep_{12}}{M_2^2}\right)^2 s \label{eq:17}
\end{eqnarray}
with $s=(p_\p + p_\g)^2$. The total cross-section for (\ref{B}) is
\begin{eqnarray}
\sigma_{\g'}= \f{5}{768\pi} \left(\f{\ep_{12}}{M_2^2}\right)^2 s \label{eq:18}
\end{eqnarray}
with $s=(p_\g + p_{\g'})^2$. In (\ref{eq:17}) and (\ref{eq:18}) all particle masses are neglected in front of $s$.
To estimate the opacity we set
$s\simeq 4\langle E_\g^2\rangle $ and $\langle E_\g^2\rangle\simeq\,
10.3\,T^2$,
where $T$ is the temperature of the medium. We get
\begin{equation}
\langle \lambda_\p\rangle\simeq \f{1}{\sigma_\p n_\g}\simeq
5\times\e{7}\ep^{-2}_{12}\left( \f{M_2}{\rm keV}\right)^4
\left( \f{T}{\rm keV}\right)^{-5} \,{\rm cm}
\label{lambdaphi}
\end{equation}
and
\begin{equation}
\langle \lambda_{\g'}\rangle\simeq \f{1}{\sigma_{\g'} n_\g}\simeq
1\times\e{6}\ep^{-2}_{12}\left( \f{M_2}{\rm keV}\right)^4
\left( \f{T}{ \rm keV}\right)^{-5} \, {\rm cm}
\label{lambdaparaph}
\end{equation}

Requiring a large enough opacity, eq.(\ref{eq:5}), for the
conditions of the Sun core, $T\simeq 1\,$keV, $\rho\simeq
100\,$g$\,$cm$^{-3}$, we are lead to
\begin{equation}
\f{\ep_{12}}{M_2^2}\gtrsim \f{10^{-3}}{{\rm keV}^2} \label{C}
\end{equation}
This condition comes from the reaction (\ref{B}); the process
(\ref{A}) gives a weaker condition.

In our model, the Sun is a copious emitter of low energy
axion-like particles and paraphotons. However,
there could be no axion-like particles reaching the Earth, because
of the decay $\p\go\g'\g'$. The lifetime of $\p$ with
energy $E_\p\sim 3T_{\rm escape}$ is
\begin{equation}
\tau_\p = 1.3 \times 10^{-7} \left( \f{m_\p}{\rm eV} \right)^{-3}
 \left(\f{M_2}{\rm keV}  \right)^2  \left(\f{E_\p}{m_\p}  \right)\,
{\rm s}
\end{equation}
$\p$ would decay
before reaching the Earth when the parameters of our model are such that
$\tau_\p<500\,$s. In this case, only paraphotons, from emission or decay,
would survive the journey from the Sun to the Earth.

Using (\ref{eq:9}) and the experimental value (\ref{eq:pvlas})
 and then imposing condition (\ref{C}) we find the allowed
 values for  $M_2$ and $\epsilon_{12}$.
There is a maximum value for the mixing $\epsilon_{12}
\lesssim  5\times10^{-7}$, and also a maximum value
for the scale $M_2\lesssim 25 \ {\rm eV}$.

Let us now discuss the cosmological constraints on the new
interactions. In the early universe, production of paraphotons
proceeds trough the reaction
\begin{equation}
\g\g\go\g'\g'\label{production}
\end{equation}
The interaction rate $\Gamma$ has to be compared to the expansion rate
$H$ of the universe to see whether the process (\ref{production}) is effective.
The calculation is similar to the one leading to (\ref{lambdaphi}) and
 (\ref{lambdaparaph}). Assuming a matter-dominated universe,
we have
\begin{equation}
\f{\Gamma}{H} \simeq \f{1}{200} \,
\left( \f{\epsilon_{12}}{10^{-7}}\right)^4 \, \left( \f{\rm eV}{M_2}\right)^4 \,
\left( \f{T}{\rm eV}\right)^{7/2}
\label{gh}
\end{equation}
Clearly, for the values of the parameters $M_2$ and $\epsilon_{12}$ discussed before and for $T>$ 1 eV, $\Gamma/H>1$ and thus a cosmic background of paraphotons will be born (when it is radiation that dominates, (\ref{gh}) has to be modified, but we reach the same conclusion). Once there is a $\g'$ population, the situation is catastrophic since the interaction (\ref{A}) is only
$\epsilon_{12}^2$-suppressed while (\ref{production}) is
$\epsilon_{12}^4$-suppressed.  As a consequence photons and paraphotons
would be in equilibrium for $T<$ 1 eV, in contradiction with the observation
of having a transparent universe for these low temperatures.

There might be other constraints on $\g-\g'$ interactions at high energies
coming from example from photon-photon interactions in accelerators.
However, here we should consider the issue that the vertex
could be subject to form factor effects. We will discuss about this topic in
the next Section.

\section{Suppression of the solar production} \label{suppression}

Let us investigate now a framework where the production in stellar cores is
considerably diminished. A first thing to notice is that  we should look at
(\ref{eq:1a}) as an effective lagrangian and consequently we should not expect it to be
valid
at arbitrarly high energies. The well studied $\pi^0\g\g$ vertex is similar
to (\ref{eq:1a}) and it is useful as a guideline.
The crucial point is that when one of the photons (or both) is off mass-shell
the effects of the $\pi^0$-photon transition form factor become manifest.

There are indeed
a variety of measurements where the transition form factor
of pseudoscalar mesons can be observed, from moderate $q^2$ up to large momentum transfer \cite{experiment}.
Let us emphasize that the appearance of a form factor is expected on general grounds. From the theoretical point of view, apart from the phenomenological VMD parameterization, one gets a form factor when using a quark-triangle model \cite{Ametller:1983ec}, when calculating in  perturbative QCD and when using some other methods \cite{pQCD}.  All these approaches are consistent among themselves and
are able to fit the data.
For example, when the $\pi^0\g\g$ vertex is described by a quark triangle loop with
off-shell photons, the explicit calculation of the diagram leads to a form
factor that can be identified with VMD provided one assigns constituent
masses to the internal up and down quarks \cite{Ametller:1983ec}.
Then, for high $q^2$ one has a suppression $M_\rho^2/q^2 \sim M_{u,d}^2/q^2 $.


These facts have encouraged us to postulate that the axion-like particle $\p$
is a confined bound-state of quark-like particles, that we will call preons in accordance with tradition. If for simplicity we consider one fermion $f$ as the only preon, $\p$ would be the $J^P=0^-$ $\bar f f$
bound state and the coupling to two photons would proceed through a triangle
loop with $f$ circulating in it. This would result in the appearance
of a form factor effect at high energies. When both photons are on-shell there is no suppression; these are the conditions in the PVLAS experiment and in the detection setup in CAST. However, in the solar medium there would be a suppression of the
$\p$ emission rate.

Let us calculate which is the required suppression $F$ in the Primakoff amplitude for having a consistent scenario. If we call $M_{\rm pvlas}$ the value in (\ref{eq:pvlas}) and $M_{\rm cast}$ the lower bound in  (\ref{eq:3}), we should have
\begin{equation}
\left[| F |^2  \f{1}{M_{\rm pvlas}^2} \right] \  \f{1}{M_{\rm pvlas}^2} \  < \
\left[  \f{1}{M_{\rm cast}^2} \right] \  \f{1}{M_{\rm cast}^2}
\label{Fcondition}
\end{equation}
where in square brackets there is the relevant factor referred to production in the Sun and outside the brackets the factor corresponding to detection in CAST. In the lhs we assume there is suppression, while in the rhs we assume none because the CAST limit is obtained assuming no form  factor suppression in the solar production. Introducing numbers we obtain
\begin{equation}
| F |   \  < \ 2 \times 10^{-9}
\label{Fbound}
\end{equation}

We now turn our attention to the theoretical prediction for $F$, that we obtain from the calculation of the preon-triangle diagram amplitude. For invariant masses $s_1$ and $s_2$ of the photons, and values of the masses of $\p$, $m_\p$, and the internal fermion $f$, $M_f$, the amplitude $F(s_1,s_2,m_\p;M_f)$ can be put in terms of dilogarithms \cite{Ametller:1983ec}. Let us comment that $F$ is in general a complex quantity and also that, as a form factor, we normalize it as
$F(0,0,m_\p;M_f)=1$.

The values for $s_1$ and $s_2$ in the solar core will be in the keV range. Indeed,
 in the interior of the Sun the Primakoff production is started by a photon of the
  thermal bath with approximately $ \omega_P^2 \simeq (0.4\, {\rm keV})^2 \simeq s_1$,  with $\omega_P$ the  plasma frequency. The virtual photon connecting the vertex to a proton (or to any charged particle) is subject to screening effects, as discussed in \cite{debye}. These effects amount to cut the momenta contributing to the Primakoff effect with the Debye-H{\"u}ckle scale $k_{DH}$, that in the solar core is
$k_{DH}^2 \simeq (9\, {\rm keV})^2 \simeq s_2$.

Provided the mass $M_f$ is much less than $s_1$ and $s_2$, we obtain a strong suppression compatible with (\ref{Fbound}). With the values of $s_{1,2}$ mentioned above and for $m_\p \lesssim 10^{-3}$ eV
(these are the values for which a coherent effect in vacuum is expected  in the PVLAS setup)
we obtain numerically
that $F$ satisfies (\ref{Fbound}) for
\begin{equation}
M_f  \lesssim  2 \times 10^{-2}\ {\rm eV}
\label{Mf}
\end{equation}
To see a bit more clearly how the suppression arises, we have verified that the exact value for $F$, in the limit $|s_2| \gg |s_1| \gg M_f \gg m_\p$ has the behaviour
\begin{equation}
|F| \sim 10^2 \, \f{(2M_f)^2}{|s_2|}
\label{assy}
\end{equation}
Thus, $M_f$ plays the role of  the cut-off energy scale of the $\p\g\g$ vertex form factor. The scale of new physics is again a low energy scale.

Let us comment that, before, we have identified $k_{DH}^2$ with $s_2$ and that it is an approximation since the $t$-channel carries other momenta. However, $ |s_2| \gtrsim
k_{DH}^2$  always, so that, at the view of (\ref{assy}), the approximation is conservative.

There is a parameter that we have not discussed, namely
the electric charge $q_f e$ of the preon $f$. With the coupling
\begin{equation}
\f{1}{2 v}\, \bar f \gamma_\mu  \gamma_5 f \, \partial^\mu \p
\label{pff}
\end{equation}
the result of the calculation of the triangle for on-shell photons is
\begin{equation}
N\, \f{q_f ^2 \, \alpha}{\pi\, v} = \f{1}{M} \simeq  \f{1}{4\times 10^{5} \, {\rm GeV}}
\label{identify}
\end{equation}
where we have already identified the result with the coupling $1/M$ in
(\ref{eq:1a}) and with the experimental value in (\ref{eq:pvlas}). Also,
we have introduce a factor $N$ corresponding to having a confining $SU(N)$
gauge group in the preon sector.

Now we have to take into account the bound on light millicharged particles \cite{milli}
coming from BBN constraints,
\begin{equation}
q_f \lesssim 2 \times 10^{-9}
\label{qf}
\end{equation}
In a paraphoton model such a small electric charge could arise naturally
\cite{Holdom}.
Together with (\ref{identify}) it implies again low energy scales
\begin{equation}
v  \lesssim 10^{-5} \, {\rm eV}
\label{qf}
\end{equation}
(We have used $N=3$).

Let us point out that the $SU(N)$ gauge group should be totally independent of the color $SU(3)_c$ standard gauge group. Otherwise, among other undesired consequences, we would have in nature hadrons with charges near $\pm (2/3)e$, $\pm (1/3)e$,  and  $\pm (4/3)e$ that would form when binding a preon or antipreon with
quarks or antiquarks, of the kind $\bar u f$, $uuf$, etc.
Also, at tree level we should take $f$ as a singlet under the standard model,
with the small electric charge arising as a higher order effect.
With all these assumptions, we think the new force and particles could have been not noticed in other experiments. Yet, there are consequences, like the existence of bound states with higher spin, as for example a state with $J=1$ that would be unstable since it would decay into $\p \g$. Of course, from a phenomenological perspective,
it would be interesting to look for signals of the preon model. From a more theoretical point of view, here we do not
attempt to build a full model, for we think it would be premature.
Rather, we have shown a possible way to evade the astrophysical limits
on axion-like particles.

\section{Conclusions}\label{conclusions}

A recent review by Raffelt \cite{Raffelt:2005mt} emphasizes that the PVLAS result
(\ref{eq:pvlas}), interpreted in terms of a new light axion-particle coupled to photons,
would lead to the Sun burning much faster that what is actually observed.
There is the pressing issue of explaining the results in a consistent model.
Of course an independent check of the results would be most welcomed;
in fact there are interesting proposals for such type of laboratory
experiment where a high sensitivity would be reached
\cite{Ringwald:2003ns}.

In this paper, we report the work we have done trying to evade the astrophysical bounds and thus accommodating a light particle coupled to two photons.
The astrophysical limits assume 1) a flux calculated with the interaction
(\ref{eq:1a},\ref{eq:1b}) in the stellar core, and 2) that the produced particles escape the star without further interaction.

Our first attemp has been trying to find a model where 2) is not true.
In our paraphoton model with (\ref{C}),
axion-like particles $\p$ are trapped in the stellar interior
and so are the paraphotons $\g'$ produced by $\p\g$ scattering.
The large opacity makes $\p$ production not a problem.
It follows that the astrophysical
limit (\ref{eq:2}), that assumes $\p$ freely escapes,  is no longer valid.
These arguments have nothing to say about
an axion-like interpretation of the PVLAS result,
because it is an earthbound laboratory experiment,
with $\p$ produced and detected in the laboratory.

The model, however, leads to photons to interact
with paraphotons so strongly that it is excluded, at least in the simple framework
we have exposed where a cosmological background density of paraphotons
emerges.
Perhaps a more elaborated model with paraphotons, or another
model with a different
strong interaction can do the job of trapping particles
in the Sun without entering in conflict with other experiments.
Let us point out that
there are also astrophysical constraints on the $\p\g\g$ coupling
coming from red giants and from SN observations.
As far as the SN is concerned, the low
value (\ref{eq:pvlas}) makes $\p$ to be trapped in the SN core
in such a way that one does not need extra interactions \cite{Masso:1997ru}.
In any case, the exercise we have presented in Section \ref{suppression}
shows that it is not trivial to evade the astrophysical constraints.

The astrophysical bound could also be evaded if the $\p\g\g$ vertex,
while fully operating at PVLAS energies, is suppressed in the
conditions of stellar interiors. In this case the condition 1) above does
not hold. We have explored the possibility
that $\p$ is a composite particle and has a form factor leading
to a suppression of the production. We have shown that this scenario is able to explain the puzzle. Our
ideas are highly speculative since they involve preons with a new
confining force and probably a miniscule electric charge,  but notice that
we have
been inspired by the pion and the $\pi^0\g\g$ vertex, that after all have the
nice property of being real. In any case, it would be crucial to look for other phenomenological consequences of the preon model. We have not found any that rules out it obviously.

Either in the case of a strong interaction leading to trapping or
of a suppression of the production,
the astrophysical bounds on axion-like particles could be evaded.
If indeed they are evaded,
there are drastic consequences for CAST, since then
the non-observation of X-rays does not imply a limit such as
(\ref{eq:3}).

Our main conclusion is that the explanation of the PVLAS result in terms of a light
particle coupled to photons
is not necessarily in contradiction with other experiments
and observations.
Let us emphasize that taking alone the PVLAS data,
if interpreted in terms of new
light particle coupled to photons, it would already
mean an interesting piece of new physics. But there
is even more. The result, taken together with the astrophysical limits
and the CAST data, means that we have to go beyond the "mere" existence
of a new pseudoscalar $\p$ coupled to photons and even more exotic physics has to be
invoked. In the scenarios of the sort we have proposed the new physics scale is at very
low energies.
 If confirmed, it would be an exciting discovery. Otherwise, if finally the models that try to evade the astrophysical constrains are shown not to be valid, the situation will be no less exciting since an alternative explanation for the PVLAS laser experiment result will be needed.

----------------------------------------------------------------------

\begin{acknowledgments}
We thank Carla Biggio, Zurab Berezhiani, Albert Bramon,
Giovanni Cantatore, and Sergei Dubovsky for useful discussions. We are specially indebted to Georg Raffelt for insightful remarks
on a first version of the paper.
We acknowledge support from the CICYT Research Project
FPA2002-00648, from the EU network on Supersymmetry and the Early
Universe (HPRN-CT-2000-00152), and from the \textit{Departament
d'Universitats, Recerca i Societat de la Informaci{\'o}} (DURSI),
Project 2001SGR00188.
\end{acknowledgments}


\begin{thebibliography}{99}
\bibitem{Raffelt:1999tx}
For a review see, G.~G.~Raffelt,
Ann.\ Rev.\ Nucl.\ Part.\ Sci.\  {\bf 49} (1999) 163
[arXiv:hep-ph/9903472].

\bibitem{flux}
K.~van Bibber, P.~M.~McIntyre, D.~E.~Morris and G.~G.~Raffelt,
Phys.\ Rev.\ D {\bf 39} (1989) 2089;
R.~J.~Creswick, F.~T.~.~Avignone, H.~A.~Farach, J.~I.~Collar, A.~O.~Gattone, S.~Nussinov and K.~Zioutas,
Phys.\ Lett.\ B {\bf 427} (1998) 235
[arXiv:hep-ph/9708210].

\bibitem{Sikivie:1983ip}
P.~Sikivie,
Phys.\ Rev.\ Lett.\  {\bf 51} (1983) 1415
[Erratum-ibid.\  {\bf 52} (1984) 695].

\bibitem{helioscope}
D.~M.~Lazarus, G.~C.~Smith, R.~Cameron, A.~C.~Melissinos, G.~Ruoso, Y.~K.~Semertzidis and F.~A.~Nezrick,
Phys.\ Rev.\ Lett.\  {\bf 69} (1992) 2333;
S.~Moriyama, M.~Minowa, T.~Namba, Y.~Inoue, Y.~Takasu and A.~Yamamoto,
Phys.\ Lett.\ B {\bf 434} (1998) 147
[arXiv:hep-ex/9805026];
Y.~Inoue, T.~Namba, S.~Moriyama, M.~Minowa, Y.~Takasu, T.~Horiuchi and A.~Yamamoto,
Phys.\ Lett.\ B {\bf 536} (2002) 18
[arXiv:astro-ph/0204388].

\bibitem{Andriamonje:2004hi}
  K.~Zioutas {\it et al.}  [CAST Collaboration],
  Phys.\ Rev.\ Lett.\  {\bf 94} (2005) 121301
  [arXiv:hep-ex/0411033].

\bibitem{pvlas}
E.~Zavattini {\it et al.},
  arXiv:hep-ex/0507107;
see also the  contributions to workshops: G. Cantatore, http://www.shef.ac.uk/physics/idm2004.html,
talk at IDM 2004, Edinburgh, England (2004) and
E.~Zavattini, talk at "XI International Workshop on Neutrino Telescopes",
Venice, Italy (2005).


\bibitem{Maiani:1986md}
  L.~Maiani, R.~Petronzio and E.~Zavattini,
  Phys.\ Lett.\ B {\bf 175}, 359 (1986).


\bibitem{Masso:1997ru}
E.~Masso and R.~Toldra,
Phys.\ Rev.\ D {\bf 55} (1997) 7967
[arXiv:hep-ph/9702275].


\bibitem{samuel}
G.~G.~Raffelt and G.~D.~Starkman,
Phys.\ Rev.\ D {\bf 40} (1989) 942;
E.~D.~Carlson and P.~Salati,
Phys.\ Lett.\ B {\bf 218} (1989) 79.



\bibitem{Okun:1982xi}
L.~B.~Okun,
Sov.\ Phys.\ JETP {\bf 56} (1982) 502
[Zh.\ Eksp.\ Teor.\ Fiz.\  {\bf 83} (1982) 892].

\bibitem{Georgi:1983sy}
H.~Georgi, P.~H.~Ginsparg and S.~L.~Glashow,
Nature {\bf 306} (1983) 765.

\bibitem{Holdom}
B.~Holdom,
Phys.\ Lett.\ B {\bf 166} (1986) 196;
Phys.\ Lett.\ B {\bf 178} (1986) 65.

\bibitem{Foot:2000vy}
R.~Foot, A.~Y.~Ignatiev and R.~R.~Volkas,
Phys.\ Lett.\ B {\bf 503} (2001) 355
[arXiv:astro-ph/0011156].

\bibitem{Dobrescu:2004wz}
B.~A.~Dobrescu,
Phys.\ Rev.\ Lett.\  {\bf 94} (2005) 151802
[arXiv:hep-ph/0411004].

\bibitem{experiment}
R.~I.~Dzhelyadin {\it et al.},
Phys.\ Lett.\ B {\bf 94}, 548 (1980)
[Sov.\ J.\ Nucl.\ Phys.\  {\bf 32}, 516.1980\ YAFIA,32,998 (1980\ YAFIA,32,998-1001.1980)];
H.~Aihara {\it et al.}  [TPC/Two Gamma Collaboration],
Mesons
  Phys.\ Rev.\ Lett.\  {\bf 64}, 172 (1990);
H.~J.~Behrend {\it et al.}  [CELLO Collaboration],
Z.\ Phys.\ C {\bf 49}, 401 (1991);
J.~Gronberg {\it et al.}  [CLEO Collaboration],
Phys.\ Rev.\ D {\bf 57}, 33 (1998)
[arXiv:hep-ex/9707031].


\bibitem{Ametller:1983ec}
A.~Bramon and E.~Masso,
  Phys.\ Lett.\ B {\bf 104} (1981) 311;
  L.~Ametller, L.~Bergstrom, A.~Bramon and E.~Masso,
  Nucl.\ Phys.\ B {\bf 228} (1983) 301.

  \bibitem{pQCD}
G.~P.~Lepage and S.~J.~Brodsky,
Phys.\ Rev.\ D {\bf 22}, 2157 (1980);
L.~Ametller, J.~Bijnens, A.~Bramon and F.~Cornet,
Phys.\ Rev.\ D {\bf 45}, 986 (1992);
A.~Z.~Dubnickova, S.~Dubnicka, G.~Pancheri and R.~Pekarik,
Nucl.\ Phys.\ Proc.\ Suppl.\  {\bf 126}, 71 (2004)
[arXiv:hep-ph/0401007];
J.~P.~B.~de Melo, T.~Frederico, E.~Pace and G.~Salme,
Phys.\ Lett.\ B {\bf 581}, 75 (2004)
[arXiv:hep-ph/0311369];
B.~W.~Xiao and B.~Q.~Ma,
Phys.\ Rev.\ D {\bf 71}, 014034 (2005)
[arXiv:hep-ph/0501160];




 \bibitem{debye}
G.~G.~Raffelt,
 Phys.\ Rev.\ D {\bf 33} (1986) 897.

\bibitem{milli}
  S.~Davidson, B.~Campbell and D.~C.~Bailey,
  Phys.\ Rev.\ D {\bf 43} (1991) 2314;
S.~Davidson, S.~Hannestad and G.~Raffelt,
JHEP {\bf 0005} (2000) 003
[arXiv:hep-ph/0001179].

\bibitem{Raffelt:2005mt}
G.~G.~Raffelt,
arXiv:hep-ph/0504152.


\bibitem{Ringwald:2003ns}
  A.~Ringwald,
  Phys.\ Lett.\ B {\bf 569}, 51 (2003)
  [arXiv:hep-ph/0306106].





\end{thebibliography}


\end{document}